\documentclass[12pt,thmsa]{article}
\usepackage{amsmath}
\usepackage{graphicx}

\input{tcilatex}

\begin{document}

\begin{center}
{\Huge Some closed form solutions to the Navier-Stokes equations}
\end{center}

\vspace{1pt}

\begin{center}
\textbf{R. M. Kiehn}

Mazan, France

rkiehn2352@aol.com

http://www.cartan.pair.com\vspace{1pt}\bigskip
\end{center}

\begin{quote}
\textbf{\vspace{1pt}Abstract: }An algorithm for generating a class of closed
form solutions to the Navier-Stokes equations is suggested, with examples.
Of particular interest are those exact solutions that exhibit intermittency,
tertiary Hopf bifurcations, flow reversal, and hysteresis.

\bigskip
\end{quote}

\section{INTRODUCTION}

The Navier-Stokes equations are notoriously difficult to solve. However,
from the viewpoint of differential topology, the Navier-Stokes equations may
be viewed as a statement of cohomology: the difference between two non-exact
1-forms is exact. Abstractly, the idea is similar to the cohomology
statement of the first law of thermodynamics.

\begin{equation}
Q-W=dU
\end{equation}
For the Navier-Stokes case, define the two inexact 1-forms in terms of the
dissipative forces

\begin{equation}
W_{D}=\mathbf{f}_{D}\bullet d\mathbf{r=}\rho \{\nu \nabla ^{2}\mathbf{V\}}%
\bullet d\mathbf{r}
\end{equation}
and in terms of the advective forces of virtual work

\begin{equation}
W_{V}=\mathbf{f}_{V}\bullet d\mathbf{r=}\rho \{\partial \mathbf{V}/\partial
t+grad(\mathbf{V\bullet V}/2)-\mathbf{V}\times curl\mathbf{V\}}\bullet d%
\mathbf{r}
\end{equation}
Then the abstract statement of cohomology, formulated as $W_{V}-W_{D}=-dP$ ,
when divided by the common function, $\rho ,$ is precisely equivalent to an
exterior differential system whose coefficients are the partial differential
equations defined as the Navier-Stokes equations,

\begin{equation}
\{\partial \mathbf{V}/\partial t+grad(\mathbf{V\bullet V}/2)-\mathbf{V}%
\times curl\mathbf{V\}}-\{\nu \nabla ^{2}\mathbf{V\}=-}grad\,P/\rho
\end{equation}
The cohomological constraint on the velocity field, $\mathbf{V}$, is such
that the kinematically defined vector, $\mathbf{f}$,

\begin{equation}
\mathbf{f}=\mathbf{f}_{V}-\mathbf{f}_{D}
\end{equation}
is a vector field that satisfies the Frobenius integrability theorem [1].
That is,

\begin{equation}
\mathbf{f\bullet \,}curl\mathbf{\,f}=0\text{ even though }\mathbf{v\bullet \,%
}curl\mathbf{\,v\neq }0.
\end{equation}
The meaning of the Frobenius criteria is that the vector $\mathbf{f}$ has a
representation in terms of only two independent functions of $\{x,y,z,t\}$.
The Navier-Stokes equations makes this statement obvious. One of these
functions has a gradient, $gradP,$ in the direction of the tangent vector to 
$\mathbf{f,}$ and the other function, $\rho ,$ is a renormalization, or
better, a reparametrization factor for the dynamical system represented by $%
\mathbf{f}$.

These observations suggest that there must exist certain constraint
relationships on the functional forms that make up the components of any
solution vector field, $\mathbf{V}$, (which usually does not satisfy the
Frobenius condition in general)\ such that the covariant kinematic vector, $%
\mathbf{f}$, is decomposable in terms of at most two functions. If such a
constraint equation can be found in terms of the component functions that
represent $\mathbf{V}$, then its solutions may be easier to deduce than the
direct solutions of the Navier-Stokes equations. For example, the constraint
relation may involve only 1 partial differential equation rather than 3. In
fact such a single constraint relation can be found by imposing a type of
symmetry condition on the system, a symmetry condition that expresses the
existence of a two dimensional (functional) representation for the vector
field,$\,\,\mathbf{f.}$ In this article attention will be focused on the two
spatial variables, $r$ and $z$, such that the solution examples will have a
certain degree of cylindrical symmetry. As these solutions involve
dissipative terms with a kinematic viscosity coefficient, $\nu ,$ they are
not necessarily equilibrium solutions of an isolated thermodynamic system.

Closed form solutions are few in number [3], but it appears that many of the
known steady-state solutions to the Navier-Stokes equations fall into the
following class of systems: Consider a variety$\,\,\{x,y,z,t\}$ with $%
r^{2}=x^{2}+y^{2}$. Consider three arbitrary functions, $\Theta (r,z)$ and $%
\Phi (r,z,t)$, and $\Lambda (r,z)$ which are defined in terms of two
independent variables spatial variables, $(r,z),$ and time. Define the flow
field, $\mathbf{V}$ in cylindrical coordinates as,

\begin{equation}
\mathbf{V}=\Lambda (r,z)\mathbf{u}_{z}+\Theta (r,z)\mathbf{u}_{r}+\Phi
(r,z,t)\mathbf{u}_{\phi }/r,
\end{equation}
where $\mathbf{u}_{\phi }$ is a unit vector in the azimuthal direction. Note
that this vector field does not necessarily satisfy the Frobenius theorem.
Note that for simplicity, the only time dependence permitted is in the
azimuthal direction.

Substitution of this format for $\mathbf{V}$ into the equation for $\mathbf{f%
}$ will yield a vector equation of the form

\begin{equation}
\mathbf{f}=\alpha (r,z)\mathbf{u}_z+\beta (r,z)\mathbf{u}_r+\gamma (r,z,t)%
\mathbf{u}_\phi /r.
\end{equation}
The Pfaffian form $W=\mathbf{f}\circ d\mathbf{r}$ will become an expression
in two variables if the azimuthal factor $\gamma (r,z,t)$ is constrained to
the value zero. In other words, a single constraint on the functions, $%
\Theta (r,z)$ and $\Phi (r,z,t)$, and $\Lambda (r,z)$, defined by the
equation $\gamma (r,z,t)=0,\,$can be used to reduce the Pfaffian form to the
expression

\begin{equation}
\alpha (r,z)dz+\beta (r,z)dr=-dP/\rho (r,z)
\end{equation}
The left hand side represents a Pfaffian form in two variables, and
therefore always admits an integrating factor. It is this idea that is used
to find new solutions to the Navier-Stokes equations. First a solution to
constraint equation is determined. Then the Cartan 1-form of total work is
computed. The 1-form is either exact, or can be made exact by an appropriate
integrating factor. If the 1-form is exact then the Pressure is obtained by
integration. It the 1-form is not exact a suitable integrating factor is
found, and that integrating factor represents a variable fluid density, $%
\rho .$ For a given choice of integrating factor, the Pressure is again
obtained by integration.

It is also useful to consider a rotating frame of reference defined by the
equation

\begin{equation}
\mathbf{\Omega }=\omega \mathbf{u}_z.
\end{equation}
It is the choice of rotational axis that defines the cylindrical symmetry.
For such rotating systems the same technique will insure that the flow
field, $\mathbf{V}$, is a solution of the Navier-Stokes equations in a
rotating frame of reference,

\begin{eqnarray}
&&\partial \mathbf{V}/\partial t+grad(\mathbf{V\circ V}/2)-\mathbf{V}\times
curl\mathbf{V} \\
&=&-gradP/\rho +\nu \nabla ^2\mathbf{V}-2\mathbf{\Omega \times V-\Omega
\times (\Omega \times r)}
\end{eqnarray}

By direct substitution, into the Navier-Stokes equation above, of the
presumed format for the velocity field $\mathbf{V}$ yields an expression for 
$\gamma (r,z)$ in terms of the three functions $\Theta (r,z)$ and $\Phi
(r,z) $, and $\Lambda (r,z):$

\begin{eqnarray}
\gamma (r,z) &=&\{\partial \Phi /\partial t+\Lambda (r,z)\partial \Phi
/\partial z+\Theta (r,z)(\partial \Phi /\partial r-2\omega r) \\
&&-\nu \{\partial ^2\Phi /\partial z^2+\partial ^2\Phi /\partial
r^2-(\partial \Phi /\partial r)/r\}.
\end{eqnarray}
Similar evaluations of the standard formulas of vector calculus in terms of
the assumed functional forms for the velocity field lead to the useful
expressions:

\begin{equation}
div\mathbf{V}=\partial \Theta /\partial r+\Theta /r+\partial \Lambda
/\partial z
\end{equation}

\begin{equation*}
curl\mathbf{\,V}=\{\partial \Phi /\partial r\,\mathbf{u}_{z}-\partial \Phi
/\partial z\,\mathbf{u}_{r}\}/r+\{\partial \Theta /\partial z-\partial
\Lambda /\partial r\}\,\mathbf{u}_{\phi }
\end{equation*}

\begin{eqnarray}
curl\,curl\mathbf{\,V} &=&\{-\partial ^2\Lambda /\partial r^2+\partial
^2\Theta /\partial z\partial r\}\,\mathbf{u}_z+ \\
&&\{-\partial ^2\Theta /\partial z^2+\partial ^2\Lambda /\partial z\partial
r\}\,\,\mathbf{u}_r+  \notag \\
&&\{-\partial ^2\Phi /\partial z^2-\partial ^2\Phi /\partial r^2+(\partial
\Phi /\partial r/r)\}\,\mathbf{u}_\phi /r  \notag
\end{eqnarray}

\begin{eqnarray}
\mathbf{V}\times curl\,\mathbf{V} &=&\Theta \{\partial \Theta /\partial
z-\partial \Lambda /\partial r\}\,\mathbf{u}_z+ \\
&&\Lambda \{\partial \Lambda /\partial r-\partial \Theta /\partial z\}\,\,%
\mathbf{u}_r+  \notag \\
&&\{(1/r^2)\,grad\,(\Phi ^2/2)-\{\Lambda \partial \Phi /\partial z+\Theta
\partial \Phi /\partial r\}\,\mathbf{u}_\phi /r  \notag
\end{eqnarray}

\begin{eqnarray}
grad(\mathbf{V\bullet V)}/2 &=&\{\Theta \partial \Theta /\partial z+\Lambda
\partial \Lambda /\partial z+\Phi (\partial \Phi /\partial z)/r^{2}\}\mathbf{%
u}_{z}+ \\
&&\{\Theta \partial \Theta /\partial r+\Lambda \partial \Lambda /\partial
r+\Phi (\partial \Phi /\partial r)/r^{2}\,\,-\Phi ^{2}/r^{3}\}\mathbf{u}_{r}
\notag
\end{eqnarray}

\begin{eqnarray}
grad(div\mathbf{V)} &=&\{\partial ^2\Theta /\partial z\partial r+(\partial
\Theta /\partial z)/r+\partial ^2\Lambda /\partial z^2\}\mathbf{u}_z+ \\
&&\{\partial \Theta ^2/\partial r^2+\partial ^2\Lambda /\partial r\partial
z+(\partial \Theta /\partial r)/r-\Theta /r^2\}\mathbf{u}_r  \notag
\end{eqnarray}

It is remarkable that many solutions to the Navier-Stokes equations then can
be found by using the following algorithm: Choose a functional form $\Phi
(r,z)$ of interest and then deduce functions $\Lambda (r,z)$ and $\Theta
(r,z)$ to satisfy the azimuthal constraint,

\begin{equation}
\gamma (r,z,t)=0.
\end{equation}

The flow field $\mathbf{V}$ so obtained is therefore a candidate solution to
the compressible, viscous, three dimensional Navier-Stokes equations for a
system with a density distribution, $\rho $ and a pressure, $P$. The
components of flow field so determined then permit the evaluation of the
coefficients of the Pfaffian form

\begin{equation}
W=\alpha (r,z)dz+\beta (r,z)dr
\end{equation}
If the expression is not a perfect differential, then use the standard
methods of ordinary differential equations to find an integrating factor, $%
\rho (r,z).$ The integrating factor represents the density distribution of
the resulting Navier-Stokes solution. The Pressure follows by integration.

This method is demonstrated in the next section for the known viscous vortex
examples reported in Lugt. In addition, several new closed form exact
solutions are generated by the technique. Among these closed form solutions
are exact solutions to the Navier Stokes equations (in a rotating frame of
reference) that exhibit the bifurcation classifications for N = 3 as given
by Langford [2]. In particular, exact, non-truncated solutions are given
that represent the trans-critical Hopf bifurcation, the saddle-node Hopf
bifurcation, and the hysteresis Hopf bifurcation. It has been long suspect
that many phenomena in hydrodynamics exhibit Hopf bifurcation; now these
exact solutions to the Navier-Stokes equation formally justify this
position, and are especially interesting for the understanding of slightly
perturbed Poiseuille flow and the onset of turbulence in a pipe.

\section{EXAMPLES}

In the following examples, the vector field specified has been used to
compute the various terms in the Navier-Stokes equations. The algebra has
been simplified by use of a symbolic computation program written in the
Maple syntax. For each example, the two vector components that make up the
work one form have been evaluated and are displayed with the solution. For
the divergence free cases, the pressure function also has been computed.
First, known solutions are exhibited, and are shown to be derived from the
above technique. Then a few new solutions are exhibited.

\subsection{Old solutions}

\subsubsection{Example 1. The Rankine Vortex}

\begin{equation}
\Phi (r,z)=a+(b+\omega )r^2,\,\,\,\,\,\,\Theta =0,\,\,\,\,\,\,\Lambda =1
\end{equation}

\begin{equation}
\mathbf{f}_{V}=\{-(a+br^{2})^{2}/r^{3}\}\mathbf{u}_{r}+\{0\}\mathbf{u}_{\phi
}/r+\{0\}\mathbf{u}_{z}
\end{equation}

\begin{equation}
\mathbf{f}_{D}=\{0\}\mathbf{u}_{r}+\{0\}\mathbf{u}_{\phi }/r+\{0\}\mathbf{u}%
_{z}
\end{equation}

This flow is a solution independent of the kinematic viscosity coefficient
(the velocity field is harmonic, as $\mathbf{f}_D$ $=0$) and therefore could
be construed as an equilibrium solution. This solution, for a and b equal to
piecewise constants, will generate the Rankine vortex.

As the flow is isochoric ($divV=0$), the steady pressure can be determined
by quadrature, and is given by the expression,

\begin{equation}
P=1/2(b^{2}r^{4}+4abr^{2}ln(r)-a^{2})/r^{2}
\end{equation}

\subsubsection{Example 2. Diffusion Balancing Advection.}

\begin{equation}
\Phi (r,z)=a+br^{2+m/\nu },\,\,\,\,\,\,\Theta (r,z)=m/r,\,\,\,\,\,\,\Lambda
=1,\,\,\,\omega =0
\end{equation}

\begin{eqnarray}
\mathbf{f}_{V} &=&\{-m^{2}-(a+br^{(2\nu +m)/\nu })^{2}/r^{3}\}\mathbf{u}_{r}+
\\
&&\{br^{(2\nu +m)/\nu }m(2\nu +m)/\nu r^{2}\}\mathbf{u}_{\phi }/r+\{0\}%
\mathbf{u}_{z}  \notag
\end{eqnarray}

\begin{equation}
\mathbf{f}_{D}=\{0\}\mathbf{u}_{r}+\{br^{(2\nu +m)/\nu }m(2\nu +m)/\nu
r^{2}\}\mathbf{u}_{\phi }/r+\{0\}\mathbf{u}_{z}
\end{equation}
In this case the Laplacian of the vector field is not zero, but the
dissipative parts exactly cancel the advective parts in the coefficient of
the azimuthal field, thereby satisfying the constraint condition. As the
functions depend only on r, the integrability (gradient) condition is
satisfied, and these solutions obey the Navier-Stokes equations for a system
of constant density. The Pressure function may be computed as

\begin{eqnarray}
P &=&(-\nu b(4a(m+\nu )+bmr^{(2\nu +m)/\nu )})r^{(2\nu +m)/\nu )}- \\
&&(m(\nu +m)(a^{2}+m^{2}))/(2m(\nu +m)r^{2})  \notag
\end{eqnarray}

The solutions are cataloged in Lugt. As these solutions explicitly involve
the kinematic viscosity, $\nu ,$ they cannot be equilibrium solutions to
isolated systems. Instead they represent steady state solutions, far from
equilibrium. A special case exists for $m/\nu =-2$.

\subsubsection{Example 3. Burger's Solution, but with Helicity and Zero
Divergence.}

\begin{equation}
\Phi (r,z)=k(1-e^{-ar^2/2\nu }),\,\,\,\,\,\,\Theta
(r,z)=-ar,\,\,\,\,\,\,\Lambda =U+2az,\,\,\,\omega =0
\end{equation}

\begin{eqnarray*}
\mathbf{f}_{V} &=&\{-(ke^{(-ar^{2}/2\nu )}+r^{2}a-k)(ke^{(-ar^{2}/2\nu
)}-r^{2}a-k)/r^{3}\}\mathbf{u}_{r}+ \\
&&\{kra^{2}/\nu \,\,e^{(-1/2ar^{2}/\nu )}\}\mathbf{u}_{\phi }/r+ \\
&&\{2(U+2az)a\}\mathbf{u}_{z}
\end{eqnarray*}

\begin{equation}
\mathbf{f}_D=\{0\}\mathbf{u}_r+\{-kra^2/\nu \,\,e^{(-1/2ar^2/\nu )}\}\mathbf{%
u}_\phi /r+\{0\}\mathbf{u}_z
\end{equation}

This solution corresponds to a modification of Burger's solution and
exhibits a 3-dimensional flow (in 2-variables) in which the diffusion is
balanced by convection to give azimuthal cancellation. The Burgers solutions
has been modified to exhibit zero divergence. This flow in a non-rotating
frame of reference exhibits a helicity.

\begin{equation}
Helicity=(U+2az)(ka/\nu )e^{(-1/2ar^{2}/\nu )}
\end{equation}

\subsection{New Solutions}

\subsubsection{Example 4. A Beltrami Type Solution}

\begin{equation}
\Phi (r,z)=r^{2}\cos (z/a),\,\,\,\,\,\,\Theta (r,z)=r\sin
(z/a),\,\,\,\,\,\,\Lambda (r,z)=2a\cos (z/a),\;\;\omega =0
\end{equation}

\begin{equation}
\mathbf{f}_{V}=\{-r\}\mathbf{u}_{r}+\{0\}\mathbf{u}_{\phi }/r+\{-4a\cos
(z/a)sin(z/a)\}\mathbf{u}_{z}
\end{equation}

\begin{equation}
\mathbf{f}_{D}=\nu /a^{2}[\{-r\,sin(z/a)\}\mathbf{u}_{r}+\{-rcos(z/a)\}%
\mathbf{u}_{\phi }/r+\{-2a\,cos(z/a)\}\mathbf{u}_{z}]
\end{equation}

This solution is a Beltrami-like solution, has zero divergence, and can be
made time harmonic by multiplying the velocity field by any function of t. \
The flow exhibits Eckman pumping and has a superficial resemblance to a
hurricane. The time independent steady flow is a strictly Beltrami $(curl\;%
\mathbf{v}=a\;\mathbf{v})$with the vorticity proportional to the velocity
field. In all cases the helicity is given by the expression,

\begin{equation}
Helicity:=(r^{2}+4a^{2}\cos (z/a)^{2})/a.
\end{equation}
The kinetic energy is a/2 times the helicity, which is a times the
enstrophy. \ The Pressure generated from the Navier Stokes equation is given
by the expression

\begin{equation}
P=1/2(r^{2}+(r^{2}(\nu /a^{2})-4\nu )sin(z/a)+4a^{2}\sin (z/a)^{2})
\end{equation}

\subsubsection{Example 5. A Saddle Node Hopf Solution}

\begin{equation}
\Phi (r,z)=\omega r^2,\,\,\,\,\,\,\Theta (r,z)=r(a+bz),\,\,\,\,\,\,\Lambda
(r,z)=U-dr^2+Bz^2
\end{equation}

The components of the advective force and dissipative force are given by the
expressions,

\begin{eqnarray}
\mathbf{f}_{V} &=&\{r(a+bz)^{2}+(U-dr^{2}+Bz^{2})rb\}\mathbf{u}_{r}+ \\
&&\{0\}\mathbf{u}_{\phi }/r+  \notag \\
&&\{-2r^{2}(a+bz)d+2(U-dr^{2}+Bz^{2})Bz\}\mathbf{u}_{z}  \notag
\end{eqnarray}
and

\begin{equation}
\mathbf{f}_{D}=\{0\}\mathbf{u}_{r}+\{0\}\mathbf{u}_{\phi }/r+\nu \{-4d+2B]\}%
\mathbf{u}_{z}
\end{equation}
The divergence of the velocity field is given by the expression:

\begin{equation}
div\,\mathbf{V}:=2\{a+(b+B)z\}
\end{equation}

The helicity of the flow depends upon the rotation, $\omega ,$

\begin{equation}
Helicity:\omega (+r^{2}b+2U-2bz^{2})
\end{equation}
but remarkably changes for finite values of $r$ and $z$, depending on mean
flow speed, $U.$

Note that when $b=0,\,B=0,\,a=0,$ the solution is equivalent to the standard
incompressible Poiseuille solution for flow down a pipe. The vector velocity
field is not harmonic, but vector Laplacian of the velocity field is a
constant.

Without these constraints, it is remarkable that the ordinary differential
equations that represent the components of the velocity field are in one to
one correspondence with the saddle node - Hopf bifurcation of Langford. That
is, the ODE,s representing the Langford format for the SN-Hopf are given by
the expressions:

\begin{eqnarray}
dz/dt &=&\Lambda (r,z)=U-dr^{2}+Bz^{2} \\
dr/dt &=&\Theta (r,z)=r(a+bz)  \notag \\
d\theta /dt &=&\omega  \notag
\end{eqnarray}

$.$This first order system which exhibits tertiary bifurcation is associated
with an exact solution of the Navier Stokes partial differential system in a
rotating frame of reference. In principle, the method also relaxes the
constraint on incompressibility, and allows a density distribution, or
integrating factor, to be computed for an exact solution to the compressible
Navier-Stokes equations which can be put into correspondence with saddle
node-Hopf bifurcation process.

This example exhibits isochoric ($divV=0$) flow for $B+b=0,a=0.$ The steady
isochoric pressure is then determined by quadrature, and is given by the
expression,

\begin{equation}
P=b(dr^{4}/2-(r^{2}-2z^{2})U-bz^{4})/2-\nu (4d-2b)z
\end{equation}
where the constant U can be interpreted as the mean flow down the pipe. Part
of the pressure is due to geometry, and part is due to the kinematic
viscosity. Note that the pressure is independent from the viscosity
coefficient when the velocity field is harmonic; e.g. when $(2d-b)=0.$ As
the vector Laplacian of the velocity field determines the dissipation in the
system, intuition would say that the harmonic solution is some form of a
limit set for the otherwise viscous flow.

\subsubsection{ Example 6. A\ Transcritical Hopf Bifurcation}

\begin{equation}
\Phi (r,z)=\omega r^2,\,\,\,\,\,\,\Theta (r,z)=r(A-a+cz),\,\,\,\,\,\,\Lambda
(r,z)=br^2+Az+Bz^2
\end{equation}

\begin{eqnarray}
\mathbf{f}_{V} &=&\{r(A-a+cz)^{2}+(br^{2}+Az+Bz^{2})rc\}\mathbf{u}_{r}+ \\
&&\{0\}\mathbf{u}_{\phi }/r+  \notag \\
&&\{2r^{2}(A-a+cz)b+(br^{2}+Az+Bz^{2})(A+2Bz)\}\mathbf{u}_{z}  \notag
\end{eqnarray}

\begin{equation}
\mathbf{f}_{D}=\{0\}\mathbf{u}_{r}+\{0\}\mathbf{u}_{\phi }/r+\nu \{4b+2B\}%
\mathbf{u}_{z}
\end{equation}
This example exhibits isochoric ($divV=0$) flow for $a=3A/2$ and $B=-c.$ The
steady isochoric pressure is then determined by quadrature, and is given by
the expression,

\begin{equation}
P=-1/4cbr^{4}-1/8A^{2}r^{2}+1/2A^{2}z^{2}-Az^{3}c+1/2c^{2}z^{4}-\nu (4b-2c)z.
\end{equation}
Again it is apparent that the pressure splits into a viscous and a
non-viscous component, and when the flow is harmonic ($2b-c=0),$ the
pressure is independent from viscosity, and there is no dissipation in the
flow.

The transcritical Hopf bifurcation is represented by the Langford system

\begin{eqnarray}
dz/dt &=&\Lambda (r,z)=br^{2}+Az+Bz^{2} \\
dr/dt &=&\Theta (r,z)=r(A-a+cz)  \notag \\
d\theta /dt &=&\omega  \notag
\end{eqnarray}

\subsubsection{ Example 7. A Hysteritic Hopf Bifurcation}

\begin{equation}
\Phi (r,z)=\omega r^2,\,\,\,\,\,\,\Theta (r,z)=r(a+bz),\,\,\,\,\,\,\Lambda
(r,z)=U-dr^2+Az+Bz^3
\end{equation}

\begin{eqnarray}
\mathbf{f}_{V} &=&\{r(a+bz)^{2}+(U-dr^{2}+Az+Az^{3})rb\}\mathbf{u}_{r}+ \\
&&\{0\}\mathbf{u}_{\phi }/r+  \notag \\
&&\{-2r^{2}(a+bz)d+(U-dr^{2}+Az+Az^{3})(A+3Az^{2})\}\mathbf{u}_{z}  \notag
\end{eqnarray}

\begin{equation}
\mathbf{f}_{D}=\{0\}\mathbf{u}_{r}+\{0\}\mathbf{u}_{\phi }/r+\nu \{-4d+6Az\}%
\mathbf{u}_{z}
\end{equation}
This system has the remarkable property that the vector Laplacian changes
sign at a position $z=2d/3A\,\,$ down stream. There is no global way of
making this solution isochoric, for the divergence is equal to

\begin{equation}
div\mathbf{V}=(A+2a)+2bz+3Az^{2}.
\end{equation}

The hysteretic Hopf bifurcation exhibits what has been called intermittency.
The Langford system is

\begin{eqnarray}
dz/dt &=&\Lambda (r,z)=U-dr^{2}+Az+Bz^{3} \\
dr/dt &=&\Theta (r,z)=r(a+bz)  \notag \\
d\theta /dt &=&\omega   \notag
\end{eqnarray}

\section{Acknowledments}

This work was supported in part by the Energy Lab at the University of
Houston in 1989, and was discussed at the Permb Conference in 1990

\section{References}

[1] FLANDERS, H. (1963) ''Differential Forms''. Academic Press, New York.

[2] LANGFORD, W. F. (1983) in ''Non-linear Dynamics and Turbulence'' Edited
by G.I. Barrenblatt, et. al. Pitman, London.

[3] LUGT, H. J. (1983) ''Vortex flow in Nature and Technology'' Wiley, New
York, p.33.

\end{document}